\documentclass{cpp2e}
\usepackage{epsfig, floatflt, rotating, latexsym, amssymb}
\shorttitle{Two-dimensional electron crystals...} 
\shortauthor{A.V.~Filinov et al.} 
\contvolum{41} 
\contyear{2001} 
\contnumber{1/2} 
\contpages{1-4} 

\title{Two-dimensional electron crystals in single and double layers}
\author{A.V.~Filinov$^{1,2}$, M.~Bonitz$^{1}$, and Yu.E.~Lozovik$^2$} 
\address{$^1$Fachbereich Physik, Universit{\"a}t Rostock 
Universit{\"a}tsplatz 3, D-18051 Rostock, Germany \\
$^2$Institute of Spectroscopy, 142090 Troitsk, Moscow Region, Russia \\ 
e-mail:~alex@ravel.mpg.uni-rostock.de}
\begin{document}
\setcounter{page}{1}
\makeheadings
\maketitle
\begin{abstract}
We present results of Monte-Carlo simulations for finite 2D 
single and bilayer systems. Strong Coulomb correlations
lead to arrangement of particles in configurations resembling a crystal lattice. 
For binary layers, there exists a particularly rich variety of lattice 
symmetries which depend on the interlayer separation $d$. We demonstrate that in 
these mesoscopic lattices there exist two fundamental types of ordering: 
radial and orientational.  
The dependence of the melting temperature on $d$ is analyzed, and a stabilization 
of the crystal compared to a single layer is found.\\
\end{abstract}

{\em Introduction.} The properties of a finite number of charged particles 
($N\sim 10$) in a single ``two-dimensional'' layer have been the subject 
of intensive threoretical and experimental investigation in the last
decade \cite{ashoori96}-\cite{filinov01}.
In particular, if at low temperature the density parameter 
$r_s$ is increased, 
transitions from a Fermi liquid to a so-called Wigner molecule~\cite{egger99} 
and, further, to a Wigner crystal have been found \cite{filinov01}. 
The interplay between the long-range Coulomb interaction and 
a shallow confinement potential plays in these systems a prominent
role making conventional effective single-particle approximations
unreliable. In this connection the predictions of both classical and
quantum Monte Carlo methods which treate N-body correlations rigorously, are of great
importance, in particular for the theoretical understanding of the
solid-liquid crossover.
Recently, a new 2D system has attracted the attention of several groups,
namely, bilayer structures. The phase diagram
of bilayer systems is far more rich compared to single layers. In particular,
the formation of Wigner lattices in electronic or ionic bilayers 
has been predicted, both by classical~\cite{peeters} and quantum-mechanical 
\cite{rap} studies and revealed the existence of distinct structural 
phases: rectangular, square, rhombic and triangular staggered lattices.
The transition between these phases takes place at specific
values of the interlayer distance $d$, when one of the phases becomes
energetically favorable and
may proceed in continuous or discontinuous manner. However, these predictions
have been made for {\em macroscopic} systems. Here, we extend the analysis
of bilayer stuctures to finite-size systems (e.g. electrons 
or ions in quantum dots and radio frequency traps, respectively). Our previous 
inverstigation of finite single layer structures \cite{filinov01} has revealed that
these mesoscopic systems have a richer phase diagram than their macroscopic 
counterpart, and that the phase diagram strongly depends on the particle number. 
In this paper, we demonstrate that different crystalline structures are stable
in different ranges of the interlayer distance. Moreover, 
we find that melting may procced in several stages,
and the crystal may be stabilized by choosing a proper value of $d$.

{\em Model and characteristic parameters.} 
We consider a system of $N$ charged particles of the same type interacting via the 
repulsive Coulomb potential located in two 2D layers which are a distance 
$d$ apart. In each layer a circular symmetric parabolic potential of strength $\omega_0$
is applied. The system is described by the hamiltonian
\begin{eqnarray}
\hat H &=&\sum\limits_{i=1}^N \frac{\hbar^2 \nabla_i^2}{2 m^*_i} +
\sum\limits_{i=1}^N \frac{m^*_i \omega_0^2 r_i^2}{2} +
\sum\limits_{l=1}^2\sum\limits^{N_l}_{i<j}\frac{e^2}{\epsilon_b |{\bf r}_{ij}|}
 + 
\frac{1}{2}\sum\limits^{N_1}_{i=1}\sum\limits^{N_2}_{j=1}
\frac{e^2}{\epsilon_b \sqrt{{\bf r}_{ij}^2+d^2}},
\label{Hamil}
\end{eqnarray}
\noindent where ${\bf r}_{ij}\equiv {\bf r}_i-{\bf r}_j$, 
$m^{*}$ and $\epsilon_b$ are the effective electron mass and background
dielectric constant, respectively. We use the following length and energy scales: 
$r_0$, given by $e^2/\epsilon_b {r_0}=m^*\omega^2 r^2_0/2$,
and $E_c$ - the average Coulomb energy, $E_c = e^2/\epsilon_b {r_0}$.
After the scaling transformations $\{r \rightarrow r/r_0, \;
E \rightarrow E/E_c, \,d \rightarrow d/r_0 \}$
the hamiltonian becomes
\begin{eqnarray}
\hat H &=&  \frac{n^2}{2} \sum\limits_{i=1}^N \nabla_i^2 +
\sum\limits_{i=1}^N r_i^2 +
\sum\limits_{l=1}^2\sum\limits^{N_l}_{i<j}\frac{1}{|{\bf r}_{ij}|}
 + 
\frac{1}{2}\sum\limits^{N_1}_{i=1}\sum\limits^{N_2}_{j=1}
\frac{1}{\sqrt{{\bf r}_{ij}^2+d^2}}, 
\label{Hamil2}
\end{eqnarray}
\noindent 
where $n\equiv \sqrt{2} \,l_0^{2}/r^2_0=(a^*_B/r_0)^{1/2}$, 
$a^*_B$  is the effective Bohr radius, and 
$l^2_0 = \hbar/(m^{*}\omega_0)$, is 
the extension of the ground state wave function of noninteracting trapped 
electrons. Further, we define, in analogy to macroscopic systems, 
$r_s\equiv r_0/a^*_B=1/n^2$ \cite{filinov01}. Finally, we introduce 
the dimensionless
 temperature $T \equiv k_B T/E_c$ which allows us to define the classical 
coupling parameter as $\Gamma \equiv 1/T$ 
\cite{bedanov94}.
To make reliable calculations in the crystal phase, where the Coulomb energy 
strongly exceeds the kinetic energy, standard methods, such as Hartree-Fock or 
density functional theory are not applicable. We, therefore, use classical and 
path integral (PIMC) Monte Carlo simulations in the classical and quantum 
regions of the phase diagram, respectively. 

{\em Phase boundary of the mesoscopic Wigner crystal in a single layer, $d=0$.}
Our simulations revealed that the structure of the clusters in the crystal phase 
strongly depends on the particle number and results from the competition of two 
ordering tendencies: for large particle numbers, a triangular lattice is energetically
favorable. In contrast, for small $N$, the particles tend to form shells. For intermediate 
particle numbers, $N>40$, only the outer electrons form shells, while the inner part
goes over to a triangular lattice structure. These configurations are remarkably stable
and are visible even outside the crystal phase. To distinguish the solid and liquid 
phases, we used the standard Lindemann criterion which is based on the analysis 
of the magnitude of the inter-particle distance fluctuations, see e.g. \cite{bedanov94}. 
Indeed, our simulations 
show that these fluctuations exhibit jumps (see Fig~\ref{fig3}.a below) which allow us 
to locate the phase 
boundary of the crystal. When leaving the crystal, the relative distance fluctuations
increase several times which results from particles exchanging their lattice sites 
or undergoing inter-shell transitions.
Interestingly, we find two distinguished crystal phases: first, a completely 
ordered state and second, a partially (only radially) ordered phase where 
shells can rotate with respect to each other. The transition between the two 
phases will be called orientational melting (``OM''), whereas the transition 
from the radially ordered state to the liquid-like state is radial melting (``RM''). 
We underline that this phase still resembles the crystal (it still exhibits shell
structure). Nevertheless, it is characterized by essentially increased inter-shell 
exchanges of particles and, therefore, this ``Wigner molecule'' phase \cite{egger99}
may not be identified with a crystal. 
\begin{floatingfigure}{8.5cm}
\vspace{0.95cm}
\hspace{-3.5cm}\centering
	\epsfig{file=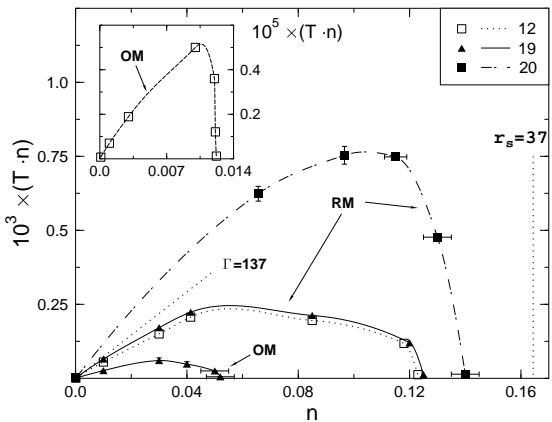, width=5.5cm}
\vspace{-1.2cm}
	\caption{\small{Phase diagram of the mesoscopic 2D Wigner crystal. ``OM'' (``RM'') denotes 
the orientational (radial) melting curves for 
$N=12,19,20$. Insert shows an enlarged picture of the 
low-density region. Dotted straight lines indicate the radial melting transition of a macroscopic 
classical and quantum  WC. Brueckner parameter follows from the density by $r_s=1/n^2$.
Shown error bars are typical for
all curves.}}
\label{fig1}
\end{floatingfigure}
\noindent The results for the phase boundaries in the density-temperature plane are summarized 
in Fig.~\ref{fig1} for various particle numbers. Consider first the line of radial 
melting ``RM''. At low densities, $n < 0.03$, we are in the classical regime, and 
the phase boundary is given by critical values of the coupling parameter 
$\Gamma$ which, in Fig.~\ref{fig1} corresponds to straight lines from the origin.
We found that $\Gamma$ deviates from the 2D bulk value,  
$\Gamma_{\infty}=137$, e.g. \cite{bedanov94}, and strongly depends on the cluster 
size. Further increase of $\Gamma$, (reduction of the slope of the straight line),
eventually leads to the second phase boundary, labelled ``OM'' which corresponds 
to freezing of the inter-shell rotation. At sufficiently low temperature, the 
electron behavior in this phase is dominated by quantum effects. This becomes 
particularly clear if the density parameter $n$ is increased. In this case, we 
observe increasing quantum (zero-point) fluctuations. When the line ``OM'' is 
reached, again a jump of the relative distance fluctuations is observed which 
corresponds to ``cold'' orientational melting \cite{filinov01}. Finally, increasing
$n$ further, leads to the line ``RM'', i.e. to cold radial melting of the 
crystal. From the figure it is clear that each of the two crystal phases exists up to a 
maximum temperature. This temperature is comparatively low. Using typical parameters of 
2D GaAs semiconductors, the maximum temperature of the radially ordered phase is in the 
range of $1\dots 5 K$.

Interestingly, the phase boundary of the crystal significantly varies with the 
number of particles. This dependence is particularly strong for the orientational 
melting transition, and we find that crystals in so-called magic clusters have 
unusually high stability. The reason is that these clusters have the highest 
angular symmetry (the particle numbers on the shells have a common divisor). This 
is a peculiarity of mesoscopie systems; the strong number dependence of the 
melting properties vanishes in the limit of a macroscopic system where also 
the two crystal phases merge into one.
 
{\em Bilayers, $d \ne 0$.} Let us now turn to electron crystallization in two parallel 
2D layers. We study a classical bilayer cystal using standard MC simulations.
The particles are arranged into two parallel layers each containing an equal number 
of particles. For zero separation, $d=0$, we have just one 2D
``atom''. For example, for $2N=38$, this is a cluster with one electron in the trap
center and three shells containing $7,13$ and $17$ electrons, respectively.
In the opposite limit, $d \rightarrow \infty$, the system consists of
two independent clusters with $N=19$ particles and shell
configuration $\{1,6,12\}$. Below we provide a detailed investigation
of the stuctural changes between these two limits for the case $N=19$. 

Different crystalline structures can be identified from snapshots of
instantaneous configurations (see insets in Fig.~\ref{fig2}). In Fig.~\ref{fig3}.c
we also show particle configurations in both layers projected onto the same plane.
To characterize the stuctural symmetry in each layer we use a suitable order
parameter given by
\begin{eqnarray}
G_{m}(R) =
 \left\langle \frac{1}{N_l} \sum^{N_l}_{j} | \phi_{m}(j) |\right \rangle ,\ \ \ 
\phi_{m}(j)= \frac{1}{N^{nb}_{j}(R)} \sum^{N^{nb}_{j}(R)}_{k} 
\exp ^{-im\theta_{jk}}
\label{order}
\end{eqnarray}
\begin{floatingfigure}{9.2cm}
\vspace{1cm}
\hspace{-4cm}\centering
	\epsfig{file=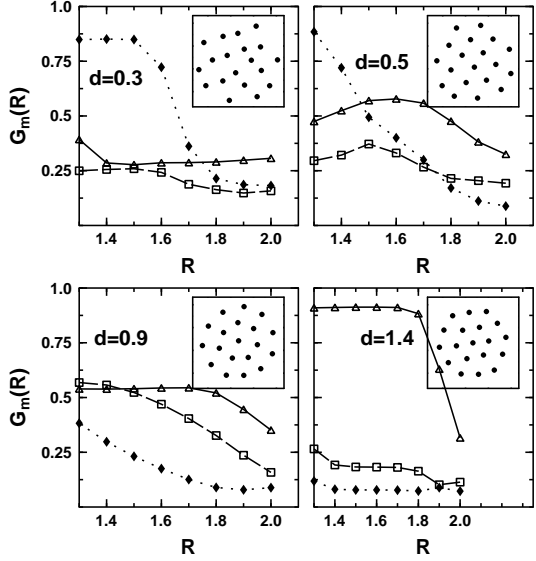, width=7cm}
\vspace{-0.5cm}
	\caption{\small{Order parameter $G_m(R)$, Eq.~(\ref{order}), 
characterizing 
the angular lattice symmetry for several values of $d$. Symbols denote
$m=4$ (diamonds), $m=5$ (squares) and $m=6$ (triangles). 
Insets show the corresponding particle configuration in one layer.
}}
\label{fig2}
\end{floatingfigure}  
\noindent where the first sum is taken over the $N_l$ particles 
in the inner part of the crystal (all particles except those on the outer shell),   
and the sum over $k$ runs over all intralayer neighbors of
particle $j$ within a circle of radius $R$. $m=2,3 \dots$;
and $\theta_{jk}$ is the angle between some fixed axis and the vector connecting
the $j-$th particle and its $k-$th neighbor. For a triangular lattice, the order 
parameter $G_6$ approaches one, whereas for a square(rhombic) lattice, $G_4$ becomes 
close to unity.
We also consider the magnitude of the relative intra/inter-layer distance fluctuations. 
In the vicinity of orientational and radial melting, 
the fluctuations, Eq.~(\ref{radial_dev}),
show a strong increase, thus providing a suitable 
quantitative criterion
 for these phase transitions (see Fig.~\ref{fig3}.a,b).
We define the interlayer ($l\ne m$, $C=N_l N_m$) and intralayer 
[$l=m$ and $i\ne j$, $C=N_l (N_l -1)$] 
distance fluctuations, $u^{lm}$, and the radial fluctuations with respect to the  
trap center, $u^l$, as
\begin{eqnarray}
u^{l m} \equiv
\frac{1}{C}\sum^{N_l}_{i} \sum^{N_m}_{j}
\sqrt{
\frac{ \langle r_{ij}^2 \rangle}{\langle r_{ij} \rangle ^2 } - 1 } 
\quad  \mbox{and}  \quad  u^l \equiv
\frac{1}{N_l}\sum^{N_l}_{i} 
\sqrt{
\frac{\langle  r_{i}^2\rangle}{\langle r_{i}\rangle ^2} - 1 }, 
\label{radial_dev} 
\end{eqnarray}
\noindent where $N_l$ and $N_m$ are the number of particles in layer $l$ and $m$,
respectively; $r_{ij}$ is the projection of the distance between particles $i$ and $j$
onto one of the layers; and $ \langle \dots \rangle $ denotes an ensemble average.

First, we start to decrease the interlayer distance from the limit 
$d \rightarrow \infty$. The configuration $\{1,6,12\}$ corresponding to two 
independent
clusters does not change up to $d \approx 0.9$. The temperature dependence
of the inter(intra)-layer pair distance fluctuations $u^{lm}(u^l)$, 
Eq. (\ref{radial_dev}), for $d=1.4$ are shown
in Fig.~\ref{fig3}.a. The two jumps in the behavior of the fluctuations
correspond to the two melting transitions. First, orientational melting
takes place in both layers simultaneously. That means that the shells with $6$ and $12$
particles are orientationally disordered and can rotate relative to each
other, as in the single-layer system. This melting takes place at a temperature 
$T_{o} \approx 2.5 \cdot 10^{-3}$. Notice that the intra and interlayer 
fluctuations differ significantly which shows that interlayer correlations are 
comparatively small which allows the shells in the two layers to rotate 
independently of each other (Fig.~\ref{fig3}.a, upper part). 
Radial melting sets in at a higher temperature, 
$T_{r} \approx 6.5 \cdot 10^{-3}$, which is the same temperature as was found for 
single-layer systems \cite{filinov01}. 
\begin{floatingfigure}{10.7cm}
\vspace{-5.4cm}
\hspace{-1.5cm}\centering
	\epsfig{file=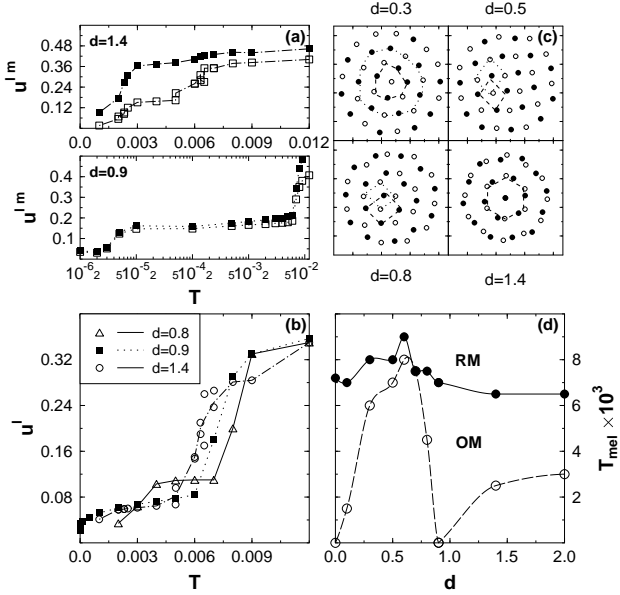, width=10.5cm}

\vspace{-1.7cm}
	\caption{\small{Crystallization in a classical bilayer system for 
varying interlayer distance $d$. Left three figures: Relative two-particle 
distance fluctuations $u^{lm}$ for particles from the same layer (open symbols) and 
from different layers (full symbols) and radial fluctuations $u^l$, Eq.~(4). 
Upper right figure shows snapshots of the crystal structure for $T=3\cdot 10^{-3}$. 
Full and open symbols denote particles from different layers, thin lines are guide 
for the eye to underline the cluster symmetry. Lower right figure 
shows the critical temperature of the radial (RM) and orientational (OM) melting 
transitions versus $d$. \vspace{0.15cm}
}}
\label{fig3}
\end{floatingfigure}
Decreasing the interlayer distance to $d=0.9$ leads to redistribution
of particles on the shells. The cluster configuration $\{1,7,11\}$ in each 
layer becomes energetically favorable, cf. Fig.~\ref{fig2}. These changes in the
cluster symmetry lead to quite different temperatures of
 orientational melting. Now the inner shells in the two layers
are ``frozen'' and cannot rotate with respect to each other. However, they can rotate
together relative to the (frozen) two outer shells which is clearly seen in the 
coinciding inter and intralayer distance fluctuations, Fig.~\ref{fig3}.a, lower part. 
This is a new type of 
disordering transition which is missing in the single-layer system. The critical 
temperature of this transition is significantly lower, $T_{o} \approx 5\cdot 10^{-6}$,
 which is a result of the modified shell configuration. Total melting 
occures at $T_{r} \approx 7 \cdot 10^{-3}$.

Now we consider the distance $d=0.8$. The interlayer correlations
lead to the formation of staggered rhombic lattices in the
inner parts of the clusters. The interesting point here is that these
staggered lattices still have the possibility to rotate relative to the
outer shells. This takes place at $T_{o} \approx 4.5 \cdot 10^{-3}$. This
essential increase of the orientational melting temperature (compared to 
the case $d=0.9$) is due to the fact that the
 outer and inner shells begin to lose 
angular symmetry (which is similar to $d=0.5$, see inset of Fig.~\ref{fig2}). 
Close inspection of this figure shows that some particles (in the top part) 
are moved inbetween the shells which effectively hampers rotation 
of the shells. The calculated 
radial distribution functions (not shown) also confirm this conclusion.
In addition to the peaks corresponding to the two shells
(for $d \ge 0.9$), here a third maximum arises inbetween.
Radial melting sets in at $T_{r} \approx 7.5 \cdot 10^{-3}$, which is an important
increase of the stability against radial disordering compared to larger values 
of $d$, cf. Fig.~\ref{fig3}.b.
  
At interparticle distances in the range 
$d=0.5 \dots 0.7$ the angular shell symmetry in the system is practically lost 
and the particle configuration in each layer resembles a rhombic 
lattice, cf. Fig.~\ref{fig3}.c. 
This leads to drastic changes in the behavior of the fluctuations: orientational
and radial transitions take place approximatly at the same temperature, 
$T_{o} \approx 7 \dots 8 \cdot 10^{-3}$
and $T_{r} \approx 7.5 \dots 9 \cdot 10^{-3}$, cf. Fig.~\ref{fig3}.d.
Further decrease of the interlayer separation up to $d=0.3$ gives rise to another
symmetry change, to formation of a rectangular lattice in each layer, 
cf. Fig.~\ref{fig2}. Interestingly however, the  
system as a whole regains spherical symmetry: as is clearly seen 
in Fig.~\ref{fig3}.c, all particles together arrange into a 
spherically symmetric cluster of $2N=38$ particles with the shell configuration 
$\{1,7,13,17\}$. This restoration of spherical symmetry causes 
the temperature of the orientational melting to drop significantly around
$d = 0.3$, cf. Fig.\ref{fig3}.d.  
The obvious conclusion is that the interlayer correlations already dominate 
the behavior of the clusters, and the system has become effectively a 
single-layer structure. When $d \rightarrow 0$, we again observe 
all properties of a single-layer 
crystal, including the critical values of temperatures for 
radial and orientational melting, $T_{o} \approx 1 \cdot 10^{-6}$ 
and $T_{r} \approx 7.2 \cdot 10^{-3}$, respectively.    


In summary, we have presented a detailed analysis of Wigner crystallization 
of finite electron systems in one and two layers. In particular, we discussed 
the influence of interlayer correlations on the crystal phase. We have shown 
that, when $d$ is reduced, the clusters first lose spherical symmetry and 
transform into a rectangular (rhombic) lattice. Finally, at low $d$, a spherical 
arrangement is re-established when all particles effectively form a single layer. 
We have found that this intermediate range of $d$-values provides an additional 
stabilization of the Wigner crystal, and the melting temperature may rise by 
up to $50\%$ (confirming the previous finding for macroscopic systems \cite{peeters}). 
The same tendency is expected to hold in mesoscopic clusters in 
the quantum range (very low temperature): interlayer correlations should significantly 
reduce the critical value of $r_s$ \cite{filinov01} up to $r_s \rightarrow 2/3 r_s$  
which should improve prospects for an experimental observation 
of electron(hole) crystallization in semiconductor heterostructures 
even at zero magnetic field.

\begin{acknowledgements}
This work is supported by the Deutsche Forschungsgemeinschaft
(Schwerpunkt ``Quantenkoh\"arenz in Halbleitern'') and the NIC J\"ulich.
\end{acknowledgements}


\begin{thebibliography}{99}

\bibitem{ashoori96} R.C.~Ashoori, Nature (London) {\bf 379}, 413 (1996);
N.B.~Zhitenev et al., Phys. Rev. Lett. {\bf 79}, 2309 (1997)

\bibitem{bedanov94} V.M.~Bedanov, and F.M.~Peeters,
Phys. Rev. B {\bf 49}, 2667 (1994), and references therein

\bibitem{lozovik} Yu.E.~Lozovik and E.A.~Rakoch,
Phys. Rev. B {\bf 57}, 1214 (1998); A.I.~Belousov and 
Yu.E.~Lozovik, JETP Lett. {\bf 68}, 858 (1998)

\bibitem{egger99} R.~Egger, W.~H\"ausler, C.H.~Mak, and H. Grabert,
Phys. Rev. Lett. {\bf 82}, 3320 (1999)

\bibitem{filinov01} A.V.~Filinov, Yu.E.~Lozovik and  M.~Bonitz,
phys. stat. sol. (b) {\bf 221}, 231 (2000);
Phys. Rev. Lett., accepted for publication (2001)

\bibitem{peeters} G.~Goldoni and F.M.~Peeters, Phys. Rev. B
{\bf 53}, 4591 (1996); I.V.~Schweigert, V.A.~Schweigert, and F.M.~Peeters,
Phys. Rev. Lett. {\bf 82}, 5293 (1999)

\bibitem{rap} F.~Rapisarda and G.~Senatore, p. 529 of
{\it Strongly Coupled Coulomb Systems}, G.~Kalman et al. (eds.), 
Plenum Press, New York, (1998)

\end{thebibliography}
\end{document}